\documentclass[aps,prl,twocolumn,reprint,showpacs,groupedaddress]{revtex4-1}
\usepackage{amsmath,amssymb,graphicx,color}
\usepackage{times}
\usepackage[table]{xcolor}

\newcommand\ket[1]{\left|\textstyle{#1}\right\rangle}
\newcommand\bra[1]{\left\langle\textstyle{#1}\right|}
\newcommand\braket[1]{\left\langle\textstyle{#1}\right\rangle}

\newcommand\down{\downarrow}
\newcommand\up{\uparrow}

\newcommand\adag{a^\dagger}

\begin{document}

\title{Quantum Phase Transition and Universal Dynamics in the Rabi model}
\author{Myung-Joong Hwang, Ricardo Puebla, Martin B. Plenio}
\affiliation{Institut f\"{u}r Theoretische Physik and IQST, Albert-Einstein-Allee 11, Universit\"{a}t Ulm, D-89069 Ulm, Germany}
\date{\today}
\begin{abstract}
We consider the Rabi Hamiltonian which exhibits a quantum phase transition (QPT) despite consisting only of a single-mode cavity field and a two-level atom. We prove QPT by deriving an exact solution in the limit where the atomic transition frequency in unit of the cavity frequency tends to infinity. The effect of a finite transition frequency is studied by analytically calculating finite-frequency scaling exponents as well as performing a numerically exact diagonalization. Going beyond this equilibrium QPT setting, we prove that the dynamics under slow quenches in the vicinity of the critical point is universal, that is, the dynamics is completely characterized by critical exponents. Our analysis demonstrates that the Kibble-Zurek mechanism can precisely predict the universal scaling of residual energy for a model without spatial degrees of freedom. Moreover, we find that the onset of the universal dynamics can be observed even with a finite transition frequency.
\end{abstract}
\maketitle

\emph{Introduction.--} Universality plays a key role for our understanding of quantum phase transitions (QPT) in interacting quantum systems~\cite{Sachdev:2011uj}. While the concept of universality is well-established in equilibrium QPT, the question to what extent the concept of universality could be extended to non-equilibrium dynamics of QPT remains largely to be explored~\cite{Polkovnikov:2011iu,Eisert:2015ka}. For a slow quench across a QPT, the closing spectral gap at a critical point leads to a breakdown of the adiabacity regardless of the quench rate. The scaling of defect formation has been shown to be entirely controlled by critical exponents and quench rate through a successful application of the Kibble-Zurek mechanism (KZM)~\cite{Kibble:1976fm,Zurek:1985ko,*Zurek:1996ik,Laguna:1997jn,delCampo:2014eu}, originally developed for classical phase transitions, to QPT in short-range interaction models~\cite{Damski:2005cr,*Damski:2008hr,Polkovnikov:2005gr,Zurek:2005cu,Dziarmaga:2005jq,*Dziarmaga:2010da,Nikoghosyan:2013wx}. However, whether this scaling holds for fully-connected models~\cite{Botet:1982ju,*Botet:1983bf}, which lack spatial degrees of freedom, such as Dicke~\cite{Dicke:1954bl} or Lipkin-Meshkov-Glick (LMG) model~\cite{Lipkin:1965bl} remains an open problem~\cite{Caneva:2008jc,Acevedo:2014eo,Klinder:2015df}.

The Dicke model considers a system of a quantized single-mode cavity field uniformly coupled to $N$ two-level atoms. It exhibits a superradiant QPT in the thermodynamic limit ($N\rightarrow\infty$)~\cite{Hepp:1973jt,Wang:1973ky,Emary:2003da,Emary:2003bd}. While tremendous efforts have been devoted to understand the QPT of the Dicke model both in and out of equilibrium~\cite{Hepp:1973jt,Wang:1973ky,Emary:2003da,Emary:2003bd,Bastidas:2012fr,Dimer:2007da,Baumann:2010js,Baumann:2011io,Bakemeier:2012ja}, a criticality of the Rabi model~\cite{Levine:2004ix,Hines:2004cn,Ashhab:2010eh,Hwang:2010jn,Bakemeier:2012ja,Ashhab:2013ke,Bishop:1996kc,*Bishop:2001ep}, the most simplified version of Dicke model with $N=1$, has been hitherto largely overlooked. Having only two constituent particles, the Rabi model is far from being in the thermodynamic limit where a QPT typically occurs; however, a ratio of the atomic transition frequency $\Omega$ to the cavity field frequency $\omega_0$ that approaches infinity, $\Omega/\omega_0\rightarrow\infty$, can play the role of a thermodynamic limit~\cite{Bakemeier:2012ja} that allows the spectral gap to be precisely closed at the critical point~\cite{Sachdev:2011uj}.

In this letter, we firstly establish the theory of equilibrium QPT of the Rabi model. At the core of our analysis is a low-energy effective Hamiltonian that is valid for $\Omega/\omega_0\gg1$ and becomes exact in the $\Omega/\omega_0\rightarrow\infty$ limit. We derive an exact solution for eigenstates, an energy spectrum, expectation values of relevant observables, and critical exponents in the $\Omega/\omega_0\rightarrow\infty$ limit. Our solution shows that there exists a critical atom-cavity coupling strength $g_c$ beyond which the $Z_2$ parity symmetry is broken and the cavity field is macroscopically occupied. Further, the effect of a finite value of $\Omega/\omega_0$ on the QPT is analyzed in the spirit of the finite-size scaling analysis. The leading order corrections to the ground state energy, the excitation energy, the average photon number, and the variance of cavity field quadratures at the critical point are derived analytically, from which \emph{finite-frequency} scaling exponents are obtained. We also perform an exact diagonalization and find an excellent agreement with analytical results. 

Our establishment of the equilibrium QPT allows us to investigate the universality in the dynamics of the Rabi model. Particularly, we are interested in quench dynamics where the system is initially prepared in the ground state and the control parameter $g$ is tuned towards to the critical point linearly in time with a quench time $\tau_q$ starting from $g=0$~\cite{Zurek:2005cu,Damski:2005cr, Polkovnikov:2005gr,Dziarmaga:2005jq, *Dziarmaga:2010da,DeGrandi:2010el, DeGrandi:2010fs, DeGrandi:2010gm}. On the one hand, we solve the dynamics exactly in the $\Omega/\omega_0\rightarrow\infty$ limit, and calculate the residual energy as a measure of the degree of non-adiabacity, which shows a power-law scaling with the quench time $\tau_q$. On the other hand, we obtain such a scaling solely from the critical exponents found in the first part of the letter. To this end, we apply KZM to the adiabatic perturbation theory~\cite{Polkovnikov:2005gr,DeGrandi:2010fs,DeGrandi:2010el,DeGrandi:2010gm} and the dynamical critical function method~\cite{Acevedo:2014eo}, independently. Both approaches give rise to the same universal scaling that precisely predicts the exact dynamics, demonstrating that the KZM can lead to a universal dynamics for a model without spatial degrees of freedom. 

Finally, we consider the same quench dynamics with a finite value of $\Omega/\omega_0$, and show that, as one decreases the ratio $\Omega/\omega_0$, there is a crossover from the universal scaling to the  $\tau_q^{-2}$ scaling, a typical scaling of the adiabatic dynamics with a finite quench time for a gapped system~\cite{Caneva:2008jc,DeGrandi:2010fs,DeGrandi:2010el}. We identify a range of quench times which leads to dynamics that closely follows the universal scaling, and show that the onset of the universal dynamics can be observed for a finite $\Omega/\omega_0$. The crossover from the universal to the  $\tau_q^{-2}$ scaling is also observed in the $\Omega/\omega_0\rightarrow\infty$ limit by ending the quench of the control parameter $g$ below the critical point. It demonstrates that the spectral gap opening due to finite $\Omega/\omega_0$ has the same effect as ending the quench below the critical point in the $\Omega/\omega_0\rightarrow\infty$ limit.

\emph{Quantum phase transition.--}
We consider the Rabi Hamiltonian~\footnote{While we have omitted the so-called $A^2$-term~\cite{Rzazewski:1975if,Vukics:2014ha}, we show in the Supplementary Materials~\cite{sup} that the Dicke-type no-go theorem holds for the Rabi model too.}
\begin{equation}
\label{rabi}
H_\textrm{Rabi}=\omega_0\adag  a+\frac{\Omega}{2}\sigma_z-\lambda( a+\adag) \sigma_x
\end{equation}
where $\sigma_{x,z}$ are Pauli matrices for a two-level atom and $a$ ($a^\dagger$) is an annihilation (creation) operator for a cavity field. The cavity field frequency is $\omega_0$, the transition frequency $\Omega$, and the coupling strength $\lambda$. We denote $\ket{\up(\down)}$ as eigenstates of $\sigma_z$, and $\ket{m}$ the eigenstate of $a^\dagger a$. The parity operator, $\Pi=e^{i\pi(a^\dagger a+\frac{1}{2}(1+\sigma_z))}$, which measures an even-odd parity of total excitation number, commutes with $H_\textrm{Rabi}$. The $Z_2$ parity symmetry has been shown to be sufficient for the model to be integrable~\cite{Braak:2011hc}; however, a lack of a closed form solution makes the approach in Ref.~\cite{Braak:2011hc} not directly applicable to investigate the QPT.

In the $\Omega/\omega_0\rightarrow\infty$ limit, we firstly find a unitary transformation, $U=\exp[\frac{\lambda}{\Omega}(a+a^\dagger)(\sigma_+-\sigma_-)]$, which makes the transformed Hamiltonian, $U^\dagger H_\textrm{Rabi} U$, free of coupling terms between spin subspaces $\mathcal{H}_{\down}$ and $\mathcal{H}_{\up}$. Upon a projection onto $\mathcal{H}_{\down}$, i.e., $H_{np}\equiv\braket{\down|U^\dagger H_\textrm{Rabi}U|\down}$, we obtain an effective low-energy Hamiltonian,
\begin{equation}
\label{Hnp}
H_{np}=\omega_0\adag  a-\frac{\omega_0g^2}{4}(a+a^\dagger)^2-\frac{\Omega}{2},
\end{equation}
where $g=2\lambda/\sqrt{\omega_0\Omega}$~\footnote{The same Hamiltonian was used in Ref.~\cite{Bakemeier:2012ja} to calculate a quantum correction to a mean-field result for the diverging variance of position quadrature of the cavity field.}. Eq.~(\ref{Hnp}) can be diagonalized to give $H_{np}=\epsilon_{np}b^\dagger b-\Omega/2$ with $\epsilon_{np}=\omega_0\sqrt{1-g^2}$, which is real only for $g\leq1$ and vanishes at $g=1$, locating the QPT. The low-energy eigenstates of $H_{\textrm{Rabi}}$ for $g\leq1$ are $\ket{\phi_{np}^m(g)}=\mathcal{S}[r_{np}(g)]\ket{m}\ket{\down}$ with $\mathcal{S}[x]=\exp[\frac{x}{2}(a^{\dagger 2}-a^2)]$ and $r_{np}(g)=-\frac{1}{4}\ln(1-g^2)$. 

The failure of Eq.~(\ref{Hnp}) for $g>1$ suggests that the number of photons occupied in the cavity field becomes proportional to $\Omega/\omega_0$ so that the higher order terms cannot be neglected, i.e., \emph{superradiance} occurs; it also suggests that $\mathcal{P}_\down$ is no longer the low-energy subspace. In order to properly capture the low-energy physics, we transform $H_\textrm{Rabi}$ of Eq.~(\ref{rabi}) by displacing the cavity field $a$, i.e., $\tilde H_\textrm{Rabi} (\pm\alpha_g)=\mathcal{D}^\dagger[\pm\alpha_g]H_\textrm{Rabi}\mathcal{D}[\pm\alpha_g]$ with $\mathcal{D}[\alpha]=e^{\alpha (\adag - a)}$ and $\alpha_g=\sqrt{\frac{\Omega}{4g^2\omega_0}(g^4-1)}$, which reads
\begin{equation}
\label{displacedrabi}
\tilde H_\textrm{Rabi} (\pm\alpha_g)=\omega_0\adag  a+\frac{\tilde\Omega}{2}\tau_{z}^\pm-\tilde\lambda( a+\adag) \tau_x^{\pm}+\omega_0\alpha_g^2
\end{equation}
where $\tau_{z}^{\pm}\equiv\ket{\up^\pm}\bra{\up^\pm}-\ket{\down^\pm}\bra{\down^\pm}=\frac{\Omega}{2\tilde\Omega}\sigma_z\pm\frac{2\lambda\alpha_g}{\tilde\Omega}\sigma_x$. Eq.~(\ref{displacedrabi}) has the same structure as Eq.~(\ref{rabi})  with rescaled frequencies $\tilde\lambda=\frac{\sqrt{\omega_0\Omega}}{2g}$ and $\tilde\Omega=g^2\Omega$.Therefore, by employing the same procedure used to derive $H_\textrm{np}$, we find an effective Hamiltonian of the Rabi Hamiltonian for $g>1$ from Eq.~(\ref{displacedrabi}),
\begin{equation}
\label{Hsp}
H_{sp}=\omega_0a^\dagger a-\frac{\omega_0}{4g^4}(a+a^\dagger)^2-\frac{\Omega}{4}(g^2+g^{-2}),
\end{equation}
whose excitation energy is found to be $\epsilon_{sp}=\omega_0\sqrt{1-g^{-4}}$, which is real for $g>1$. Note that two independent choices of $\alpha=\pm\alpha_g$ in Eq.~(\ref{displacedrabi}) lead to an identical spectrum. The low-energy eigenstates of $H_\textrm{Rabi}$ for $g>1$, $\ket{\phi_{sp}^m(g)}_\pm=\mathcal{D}[\pm\alpha_g]\mathcal{S}[r_{sp}(g)]\ket{m}\ket{\down^\pm}$ where $r_{sp}(g)=-\frac{1}{4}\ln(1-g^{-4})$, are, therefore, degenerate; they also have a spontaneously broken parity symmetry, as evident from the non-zero coherence of the field $\braket{a}=\pm\alpha_g$. The higher order corrections in Eq.~(\ref{Hnp}) and (\ref{Hsp}) vanish exactly in the $\Omega/\omega_0\rightarrow\infty$ limit. Therefore, $H_{np}$ and $H_{sp}$ are the \emph{exact} low-energy effective Hamiltonian for the \emph{normal phase} ($g<1$) and $\emph{superradiant phase}$ ($g>1$), respectively, for which the subscripts $np$ and $sp$ stand. See Ref.~\cite{sup} for a detailed derivation of the effective Hamiltonian and its solution.
\begin{figure}
\centering
\includegraphics[width=1\linewidth,angle=0]{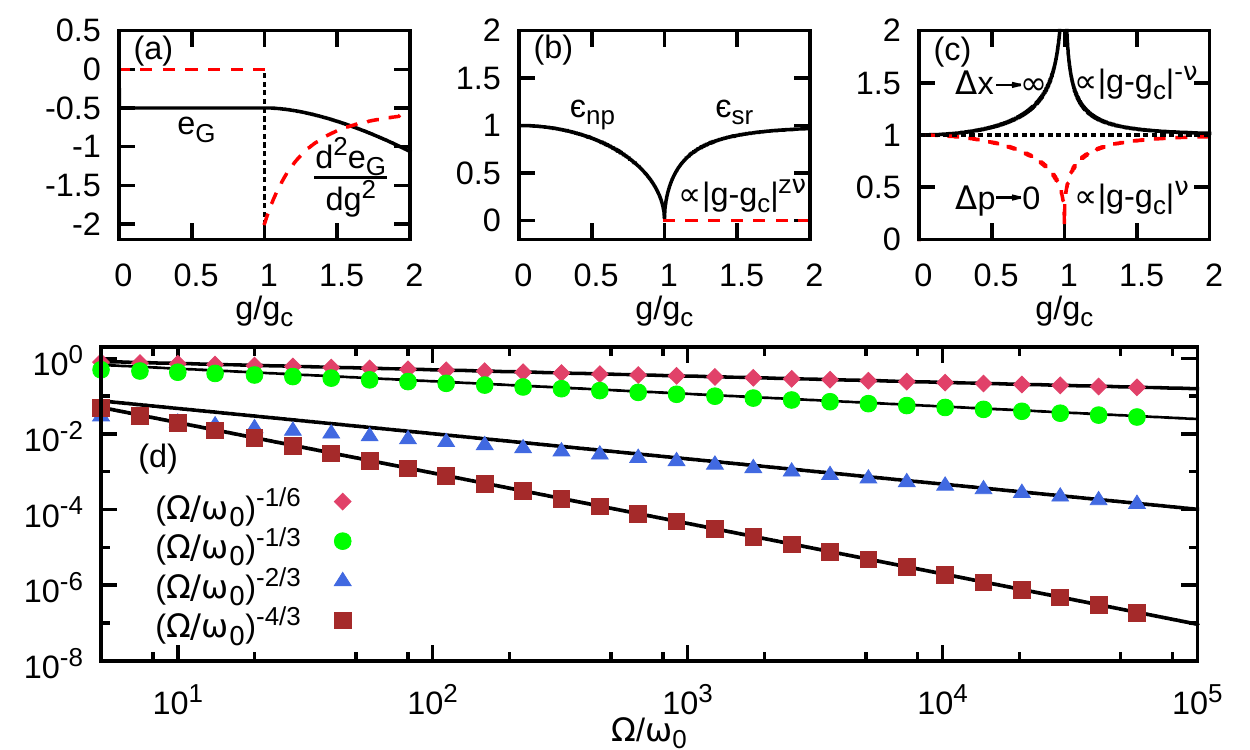}
\caption{Top panel: exact solutions of the Rabi model in the $\Omega/\omega_0\rightarrow\infty$ limit as a function of the dimensionless coupling strength $g/g_c$ for (a) the rescaled ground state energy $e_G$ (solid) and $d^2 e_G/d^2 g$ (red-dashed), (b) the excitation energy $\epsilon$ (solid) and the energy difference between the ground and the first excited state (red-dashed) showing the ground state degeneracy for $g/g_c\geq1$, and (c) the variance of position $\Delta x$ (solid) and momentum $\Delta p$ (red-dashed) quadrature of the cavity field, and $\Delta x\Delta p$ (dotted). In (b) and (c), the scaling relation near the critical point is indicated. Bottom panel: a leading order correction for finite $\Omega/\omega_0$ at $g=g_c$ for $\Delta p$, $\epsilon$, the order parameter $n_c$, and $e_G$ from top to bottom, respectively. The analytical results (lines) predict precisely the exact diagonalization results (points) for all observables. The finite-frequency scaling exponents for each observable are indicated. }
\label{fig:1}. 
\end{figure}

Our exact solution shows that the superradiant QPT occurs at the critical point $g_c=1$. The rescaled cavity photon number $n_c=\frac{\omega_0}{\Omega}\braket{a^\dagger a}$ is zero for $g< g_c$ and $n_c=(g^4-g_c^4)/4g^2$ for $g>g_c$; thus, $n_c$ is an order parameter. The rescaled ground state energy, $e_G(g)\equiv\frac{\omega_0}{\Omega}E_G(g)$ is $-\omega_0/2$ for $g< g_c$ and $-\omega_0(g^2+g^{-2})/4$ for $g>g_c$. While $e_G(g)$ is continuous, $d^2e_G(g)/d^2g$ is discontinuous at $g=g_c$, revealing the second-order nature of the QPT [Fig.~\ref{fig:1} (a)].  Near the critical point, the excitation energy in both phases, $\epsilon_{np}$ and $\epsilon_{sp}$, vanishes as $\epsilon(g)\propto|g-g_c|^{z\nu}$ with $z\nu=1/2$  [Fig.~\ref{fig:1} (b)], where $\nu$ ($z$) is the (dynamical) critical exponent. Meanwhile, the variance of position quadrature of the field $x=a+a^\dagger$ diverges as $\Delta x(g)\propto|g-g_c|^{-1/4}\propto\epsilon^{-1/2}$, from which we find that $z=2$ and $\nu=1/4$ [Fig.~\ref{fig:1} (c)]. While we have defined the critical exponents $z$ and $\nu$ separately by noticing that $\Delta x$ plays an analogous role of the diverging length scale in extended quantum systems~\cite{Sachdev:2011uj}, only is the product $z\nu$ an important exponent in the following analysis. The critical point also accompanies an infinite amount of squeezing in the momentum quadrature $p=i(a^\dagger-a)$ so that it remains in the minimum uncertainty state for any $g$, i.e., $\Delta x(g) \Delta p(g)=1$  [Fig.~\ref{fig:1} (c)].

\emph{Finite-frequency scaling.--} We complete our study of the equilibrium QPT by investigating the \emph{finite-frequency} effect. Firstly, we derive a leading order correction to the exact effective Hamiltonian. To this end, we find a unitary transformation
$U^\Omega=\exp[(\frac{\lambda}{\Omega}(a+a^\dagger)-\frac{4\lambda^3}{3\Omega^3}(a+a^\dagger)^3)(\sigma_+-\sigma_-)]$ of Eq.~(\ref{rabi}) that decouples the $\mathcal{H}_{\down}$ and $\mathcal{H}_{\up}$ subspaces up to \emph{fourth} order in $\lambda/\Omega$ and project to $\mathcal{H}_{\down}$ to obtain~\cite{sup}
\begin{equation}
\label{HnpOmega}
H^\Omega_{np}=H_{np}+\frac{g^4\omega_0^2}{16\Omega}(a+a^\dagger)^4+\frac{g^2\omega_0^2}{4\Omega},
\end{equation}
where the leading order correction adds a quartic potential for the cavity field. Although $H^\Omega_{np}$ is not exactly solvable, a variational method can be used to derive analytical expectation values~\cite{sup}. We find that, at the critical point, the excitation energy vanishes and the characteristic length scale diverges with a power-law scaling,
\begin{align}
\label{finitescaling}
\epsilon_{g_c}(\Omega/\omega_0)=\omega_0\left(\frac{2\Omega}{3\omega_0}\right)^{-1/3},~
\Delta x_{g_c}(\Omega/\omega_0)=\left(\frac{2\Omega}{3\omega_0}\right)^{1/6}.
\end{align}
In addition, the leading order correction for $e_G$ and $n_c$ are given by $e_{G,g_c}(\Omega/\omega_0)=(\omega_0/4)(2\Omega/3\omega_0)^{-4/3}$ and $n_{c,g_c}(\Omega/\omega_0)=1/6(2\Omega/3\omega_0)^{-2/3}$.  The exponents of these scaling relations, the finite-frequency scaling exponents, are found to be the  same as the finite-size scaling exponents of corresponding observable for the Dicke model~\cite{Vidal:2006ex} and LMG model~\cite{Dusuel:2004eo,Dusuel:2005hw}, which also have the same critical exponent $z$ and $\nu$~\cite{Gilmore:1986gf,Emary:2005cr}. We perform an exact diagonalization of Eq.~(\ref{rabi}) and show that the numerically obtained scaling exponents precisely match the analytical results [Fig.~\ref{fig:1} (d)]. 

\begin{figure}
\centering
\includegraphics[width=1\linewidth,angle=0]{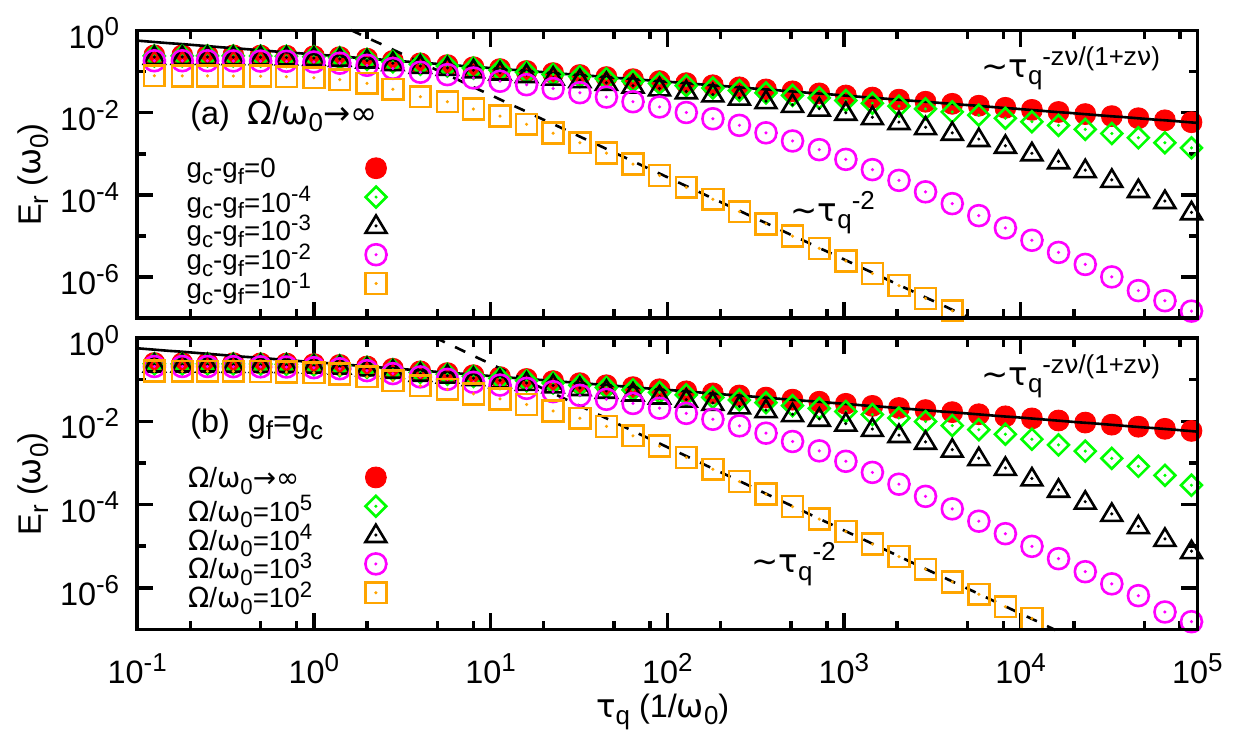}
\caption{Residual energy $E_r$ as a function of the quench time $\tau_q$ obtained by solving a nearly adiabatic dynamics for (a) different values of the final coupling strength $g_f$ ranging from $g_f=0.9g_c$ to $g_f=g_c$ (from bottom to top) in the $\Omega/\omega_0\rightarrow\infty$ limit and (b) different ratios $\Omega/\omega_0$ ranging from $\Omega/\omega_0=10^2$ to $\Omega/\omega_0\rightarrow\infty$  (from bottom to top) with a fixed final coupling strength $g_f=g_c$. For  $g_f=g_c$  in the $\Omega/\omega_0\rightarrow\infty$ limit, it precisely follows the universal scaling relation (solid line), predicted by the Kibble-Zurek mechanism.  Both moving $g_f$ away from $g_c$ and reducing the ratio $\Omega/\omega_0$ result in a crossover from the universal scaling to $\tau_q^{-2}$ scaling (dashed line).}
\label{fig:2}
\end{figure}

\emph{Universal scaling for adiabatic dynamics.--} Having established the equilibrium QPT of the model, we are now able to investigate the dynamics of the QPT. We consider a protocol where the control parameter $g$ is changed linearly in time, $g(t)=g_ft/\tau_q$, with $g_f$ being the final value. The system is initially in the ground state. As $g(t)$ approaches the critical point, the vanishing spectral gap makes the relaxation time of the system diverge, inevitably creating quasiparticle excitations irrespective of how large the quench time $\tau_q$ is. Applying KZM~\cite{Polkovnikov:2011iu,Kibble:1976fm,Zurek:1985ko,*Zurek:1996ik,Laguna:1997jn,Damski:2005cr,Polkovnikov:2005gr,Zurek:2005cu,Dziarmaga:2005jq,*Dziarmaga:2010da,delCampo:2014eu,Damski:2008hr}, we define a time instant $\hat t$ that divides the dynamics into the adiabatic and impulsive regime from $\eta^2(t)=\dot \eta(t)$, where the accessible energy gap $\eta$ is given as $\eta=2\epsilon_{np}$ for $g<g_c$ due to the parity symmetry. From $\epsilon_{np}=\omega_0\sqrt{1-g^2}$, we find $\hat{g}\sim g_c-(4\sqrt{2}\omega_0\tau_q)^{-\frac{1}{z\nu+1}}$~\cite{sup}  where the coupling instant $\hat g\equiv g(t=\hat t)$ moves away from the critical point as one decreases the quench time so that the impulsive regime widens. Note that we only consider $g(t)\leq g_c$ for a simplicity~\cite{DeGrandi:2010el,DeGrandi:2010gm,DeGrandi:2010fs}.

The wave function at time $t$ can be expressed in terms of the instantaneous eigenstates of $H_{\textrm{np}}(g(t))$, i.e., $\ket{\Psi(t)}=\sum_m c_m(t)\mathcal{S}[r_{np}(t)]\ket{m}$. Then, we apply the adiabatic perturbation theory~\cite{Polkovnikov:2005gr,DeGrandi:2010el,DeGrandi:2010gm} to calculate the residual energy $E_r$ at the end of the quench, which measures the degree of non-adiabacity, defined as $E_r\equiv\braket{\Psi(\tau_q)|H_{np}(g(\tau_q))|\Psi(\tau_q)}-E_G(g(\tau_q))$. For a protocol that stays in the adiabatic regime, i.e., $g_f\ll\hat g$, we obtain a scaling relation, $E_r\propto\tau_q^{-2}$~\cite{sup}, which is a typical scaling for the adiabatic dynamics with a finite quench time for a gapped Hamiltonian. If the protocol involves the impulsive regime, $g_f\sim \hat g$, we find that the residual energy follows a universal scaling relation,
\begin{equation}
\label{heat}
E_r\propto\tau_q^{-z\nu/(z\nu+1)},
\end{equation}
that is, $E_r\propto\tau_q^{-1/3}$ since $z\nu=1/2$~\cite{sup}. A different way to predict the universal scaling of $E_r$ based on KZM is to use the dynamical scaling function approach~\cite{Acevedo:2014eo}, which expresses the scaling relation in terms of the finite-frequency scaling exponents. We confirm that it predicts the same universal scaling relation as in Eq.~(\ref{heat})~\cite{sup}.

For short-range interaction models, the residual energy due to a slow quench stems from spatial defects in order parameter across a QPT, whose scaling has been successfully predicted by KZM~\cite{Zurek:2005cu,Dziarmaga:2005jq,*Dziarmaga:2010da,delCampo:2014eu}. However, it is not clear whether KZM can predict the scaling of the residual energy in fully-connected models due to their lack of spatial degrees of freedom. In fact, although the same scaling relation with Eq.~(\ref{heat}) has also been predicted for the Dicke and LMG model~\cite{Acevedo:2014eo}, a numerical calculation with a finite-size LMG model shows a significant discrepancy with the universal scaling as it estimates $E_r\propto\tau_q^{-3/2}$~\cite{Caneva:2008jc}, raising a doubt on the applicability of the KZM to the fully-connected models~\cite{Acevedo:2014eo}. Strictly speaking, one has to solve the dynamics exactly in the thermodynamic limit for the LMG or Dicke model, or equivalently in the $\Omega/\omega_0\rightarrow\infty$ limit for the Rabi model to test the validity of the universal scaling relation, which is accomplished in the following section.

\emph{Exact solution for adiabatic dynamics.--} The exact low-energy effective Hamiltonian given in Eq.~(\ref{Hnp}) allows one to numerically solve the slow quench dynamics of the Rabi model, which involves only a small number of quasiparticle excitations, in the $\Omega/\omega_0\rightarrow\infty$ limit. The equation of motion is given as $i\dot a_H(t)=[a_H(t),H_{np,H}(t)]$, where the subscript $H$ indicates the operators in the Heisenberg picture. We express the cavity field operator at time $t$ as  $a_H(t)=u(t)a(0)+v^*(t)a^\dagger(0)$ with an initial condition $u(0)=1$ and $v(0)=0$, and  $|u(t)|^2-|v(t)|^2=1$, and derive coupled differential equations for $u(t)$ and $v(t)$,
\begin{align}
\label{eom}
\frac{i}{\omega_0}\frac{d u(t)}{d t}&=\left(1-\frac{g^2(t)}{2}\right)u(t)-\frac{g^2(t)}{2}v(t),\nonumber\\
-\frac{i}{\omega_0}\frac{d v(t)}{d t}&=\left(1-\frac{g^2(t)}{2}\right)v(t)-\frac{g^2(t)}{2}u(t).
\end{align}
The residual energy in terms of $u(t)$ and $v(t)$ is given by
\begin{align}
\label{residual}
E_r&=\omega_0|v(t)|^2-\frac{\omega_0g_f^2}{4}|u(t)+v(t)|^2-\frac{\epsilon_{np}(g_f)-\omega_0}{2}.
\end{align}
\begin{figure}
\centering
\includegraphics[width=1\linewidth,angle=0]{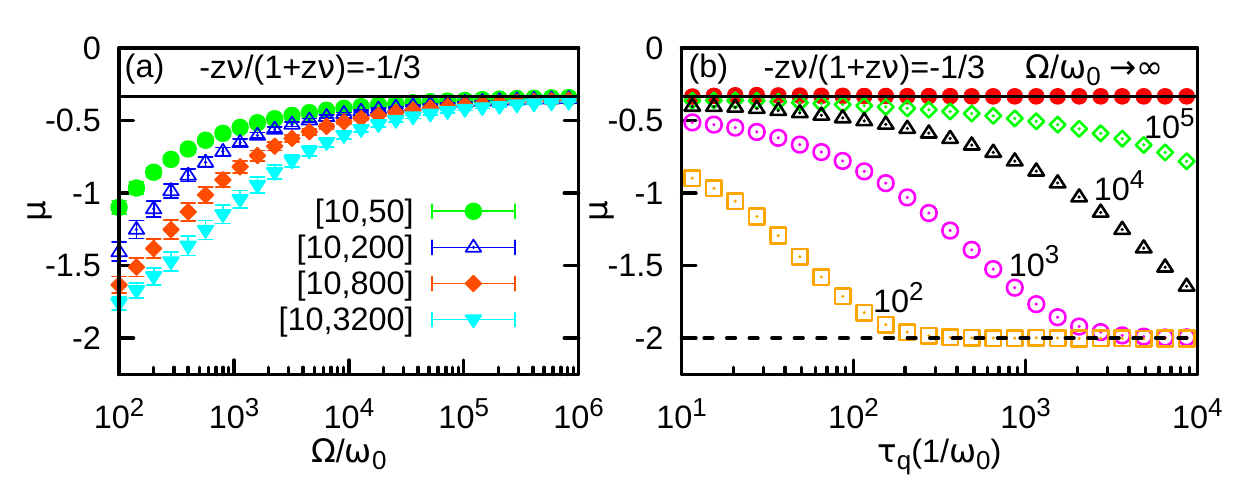}
\caption{Exponents of power-law scaling  $\tau_q^{\mu}$ for the residual energy $E_r$ presented in Fig.~\ref{fig:2} (b). (a) Fits obtained for different ranges of the quench time $[\tau_{q1},\tau_{q2}]$. The values for $\tau_{q1(q2)}$ are indicated in the figure. The exponent $\mu$ converges to the universal scaling exponent $-1/3$ for finite $\Omega/\omega_0$. (b) Fits obtained for a range of the quench time $[\tau_q\Delta\tau_q,\tau_q/\Delta\tau_q]$ as a function of $\tau_q$ with a fixed log-scale interval $\log_{10}\Delta\tau_q=6.25\times10^{-2}$. The crossover from $\mu=-2$ to $\mu=-1/3$ as one increases $\Omega/\omega_0$ is clearly demonstrated.}
\label{fig:3}. 
\end{figure}
In Fig.~\ref{fig:2} (a), we plot $E_r$ at the end of the quench as a function of $\tau_q$ for different values of the final coupling strength $g_f$ in the $\Omega/\omega_0\rightarrow\infty$ limit. For a protocol that ends right at the critical point, $g_f=g_c$, it precisely follows the universal scaling given in Eq.~(\ref{heat}).  It confirms that the nearly adiabatic dynamics of the QPT in the Rabi model can be completely characterized by the critical exponents alone, thus is universal. We note that the saturation of $E_r$ observed for a short quench time, $\tau_q\lesssim1/\omega_0$, corresponds to sudden quench dynamics. As we change $g_f$ progressively away from the critical point, $g_f<g_c$, the universal scaling breaks down and the $\tau_q^{-2}$ scaling emerges, which is precisely the scaling predicted using the adiabatic perturbation theory in the adiabatic regime~\cite{sup}. 

We find that the leading order correction to the equation of motion for finite $\Omega/\omega_0$ adds 
an additional term, $f(u,v)=(3\omega_0/4\Omega) g^4(t)(u+v)|u+v|^2$, to the right hand side of both equations in Eq.~(\ref{eom})~\cite{sup}. For a quench that ends at the critical point $g_f=g_c$, the leading order correction to the residual energy adds an additional term, $h(u,v)=(3\omega_0^2g_c^4)/(16\Omega)|u+v|^4-\frac{\omega_0}{4}
{(2\Omega/3\omega_0)}^{-1/3}$, to Eq.~(\ref{residual})~\cite{sup}. In Fig.~\ref{fig:2} (b),  where we reduce the ratio $\Omega/\omega_0$ from infinity to $10^2$ for $g_f=g_c$, we observe a crossover behavior for the residual energy virtually identical to Fig.~\ref{fig:2} (a). This is because a finite value of $\Omega/\omega_0$ opens up an energy gap at $g_f=g_c$ whose effect is equivalent to ending  the protocol away from the critical point. 

An interesting aspect of the crossover behavior for the scaling of $E_r$ shown in Fig.~\ref{fig:2} (b) is that there is a range of quench time $\tau_q$ at around $\tau_q\in[10,10^3]$ where the $E_r$ closely follows the universal power-law even for finite values of $\Omega/\omega_0$. By closer inspection, we find fits for the slope of curves in Fig.~\ref{fig:2} (b), which corresponds to the exponents of power-law scaling of $E_r$, for a wide range of quench times. As shown in Fig.~\ref{fig:3} (a), the exponents converge to the universal scaling exponent $-1/3$ as one increases the ratio $\Omega/\omega_0$, showing that the onset of the universal dynamics can be observed with finite $\Omega/\omega_0$. The convergence to the universal scaling implies that the energy gap whose scaling is given in Eq.~(\ref{finitescaling}) is sufficiently small to drive the system into the impulsive regime so that the dynamics is strongly influenced by the nature of the critical point. As the energy gap widens for smaller values of $\Omega/\omega_0$, the influence of the critical points gradually vanishes, leading to a crossover to $\tau_q^{-2}$ scaling. In Fig.~\ref{fig:3} (b), the crossover of the scaling from $\tau_q^{-1/3}$ to $\tau_q^{-2}$ is further elucidated by finding fits for much shorter interval of $\tau_q$, which approximates the slope of the tangent line of graphs in Fig.~\ref{fig:3} (b). Varying $g_f$ in the $\Omega/\omega_0\rightarrow\infty$ limit shows identical features shown in Fig.~\ref{fig:3}~\cite{sup}.

\emph{Conclusion.--} We have found an effective low-energy description of the Rabi model that unveils the universality of the model both in and out of equilibrium. Our analysis shows that the superradiant QPT which has been primarily studied for systems of thermodynamically many atoms can as well be investigated with systems of a single atom. An important advantage of the reduced degrees of freedom is that solving the critical dynamics is more tractable; indeed, we have been able to report a first confirmation of the KZM prediction for a model without spatial degrees of freedom. Together with an impressive ongoing progress of technologies to realize the interaction between a two-level system and a single harmonic oscillator, we expect that the Rabi model can serve as an excellent platform to study equilibrium and non-equilibrium critical phenomena.

\emph{Acknowledgements.--} This work was supported by the EU Integrating projects SIQS and DIADEMS, the EU STREP EQUAM and an Alexander von Humboldt Professorship.

\bibliographystyle{apsrev4-1}
\bibliography{paper}

\end{document}